# Intensity correlation OCT – a true classical equivalent of quantum OCT able to achieve up to 2-fold resolution improvement in standard OCT images

**Sylwia M. Kolenderska[1]* and Piotr Kolenderski[2]**

[1]The Dodd-Walls Centre for Photonic and Quantum Technologies, Department of Physics, University of Auckland, Auckland 1010, New Zealand

[2]Faculty of Physics, Astronomy and Informatics, Nicolaus Copernicus University, Grudziądzka 5, 87-100 Toruń, Poland

*skol745@aucklanduni.ac.nz

## ABSTRACT

Quantum Optical Coherence Tomography (Q-OCT) uses quantum properties of light to provide several advantages over its classical counterpart, OCT: it achieves a twice better axial resolution with the same spectral bandwidth and it is immune to even orders of dispersion. Since these features are very sought-after in OCT imaging, many hardware and software techniques have been created to mimic the quantum behaviour of light and achieve these features using traditional OCT systems. The most recent, purely algorithmic scheme – an improved version of Intensity Correlation Spectral Domain OCT named ICA-SD-OCT showed even-order dispersion cancellation and reduction of artefacts. The true capabilities of this method were unfortunately severely undermined, both in terms of its relation to Q-OCT and in terms of its main performance parameters. In this work, we provide experimental demonstrations as well as numerical and analytical arguments to show that ICA-SD-OCT is a true classical equivalent of Q-OCT, more specifically its Fourier domain version, and therefore it enables a true two-fold axial resolution improvement. We believe that clarification of all the misconceptions about this very promising algorithm will highlight the great value of this method for OCT and consequently lead to its practical applications for resolution- and quality-enhanced OCT imaging.

## Introduction

Optical Coherence Tomography (OCT) has become an indispensable tool in medicine[1] due to its ability to visualise internal structures of biomedical objects on a micrometre scale and in a non-contact and non-invasive way. OCT is based on an interferometric measurement of the time of flight of light and can be performed by axially translating a mirror in a reference arm as in Time-domain OCT or by keeping the mirror fixed and measuring the spectrum of light as in Fourier-domain OCT. OCT's axial resolution – so the resolution in the direction of light propagation in an object – is inversely proportional to the spectral bandwidth of the light source. Assuming a Gaussian profile, axial resolution, $\delta z$, is given by $\delta z = \frac{2\ln 2}{\pi} \frac{1}{\Delta k}$, which is traditionally[2] rewritten to $\delta z = \frac{2\ln 2}{\pi} \frac{\lambda_c}{\Delta \lambda}$ for convenience, where $\Delta k$ and $\Delta \lambda$ is the spectral bandwidth in wavenumber and wavelength, and $\lambda_c$ is the central wavelength. At first glance, it might seem that broader spectral bandwidths lead to better axial resolution. Unfortunately, the broader the spectral range is, the more prominent chromatic dispersion effects become. Whereas it is fairly easy to compensate the dispersion mismatch in the arms of an interferometer[2–8], the resolution degradation for deeper layers of an object caused by a dispersive character of these layers is almost impossible to mitigate. It was shown that the practical limit for axial resolution is 1 $\mu m$[9] and this limit has already been achieved in OCT in both visible[10] and NIR[11,12] wavelength ranges.

Because of this trade-off, further development of OCT in terms of resolution seems no longer possible through traditional means. It rather necessitates novel and unconventional approaches which go beyond the standard viewpoints and which should borrow and adapt solutions from other fields of optics. With such an attitude, inspired by MRI, Ling et al.[13] showed the first successful alternative algorithm to Fourier Transformation with which they obtained a several-fold resolution increase. Similarly, Abouraddy et al.[14] reached to quantum optics to realize the first-ever Quantum OCT (Q-OCT) – not only did this method intrinsically provide a two-fold resolution increase, but also it is free of even-order dispersion effects, which contribute most to the resolution degradation.

In Q-OCT, entangled photon pairs are created in a nonlinear crystal: one photon penetrates the object placed in one arm and the other photon is reflected from a mirror in the other arm of the interferometer. These photons overlap at a beamsplitter and the coincidence of their simultaneous arrival at photodiodes located at two output ports is measured. In the first implementation of this technique, Time-domain Q-OCT (Td-Q-OCT)[15], a depth profile of the object – an A-scan – is obtained by axially translating the reference mirror and performing the coincidence rate measurement. In Fourier-domain Q-OCT (Fd-Q-OCT)[16], the mirror is fixed and the coincidence measurement is done together with wavelength discrimination producing a two-dimensional joint spectrum. An A-scan is obtained by Fourier transforming the main diagonal of the joint spectrum. Unfortunately, both Q-OCT modalities require sources of entangled photon pairs, which are inefficient and therefore, pose significant experimental challenges. Due to low intensity levels, data acquisition time extends from at least tens of minutes to even hours and the imaging itself can only be performed for simple highly reflective objects, which excludes biomedical samples.

Nevertheless, Q-OCT was able to inspire new solutions in traditional OCT and led to the creation of quantum-mimic OCT methods aimed at recreating the features of Q-OCT using different hardware and software techniques[17–30]. With time, many misconceptions arose around the performance of these methods, including whether the axial resolution improvement is on the order of $\sqrt{2}$ or 2, or even claiming that the resolution enhancement is not present at all[29].

In the onslaught of different approaches and different points of view on the subject, the paper by Jensen *et al.*[29] has passed unnoticed and remains unappreciated – oddly, even by its own authors – despite its unquestionable relevance to Q-OCT and the potential it holds for the advancement of OCT. Here, we show that the method of Jensen *et al.* called intensity correlation spectral domain OCT (ICA-SD-OCT) is a true classical equivalent of Q-OCT. We prove that this approach is able to recreate the signal of Fd-Q-OCT and all that such signal entails: dispersion cancellation, artefacts and most importantly – what was tried to be debunked by Jensen *et al.* – almost two-fold axial resolution improvement. We provide an easy explanation of why the resolution increases by 2 in Q-OCT, how the factor of 2 can be achieved in quantum-mimic methods and finally, why the two-fold improvement is true and not merely illusory.

## Theoretical results

### General comparison of signals in Q-OCT and ICA-SD-OCT

The signal acquired in Fourier-domain Q-OCT (Fd-Q-OCT)[16] is a two-dimensional joint spectrum and can be expressed in the following mathematical form:

$$C_{\text{Fd–Q–OCT}}(\omega_1, \omega_2) = |\phi(\omega_1, \omega_2)|^2 (|f(\omega_1)|^2 + |f(\omega_2)|^2 - 2 \, \text{Re}\{f(\omega_1)f^*(\omega_2)\}) \tag{1}$$

where $\omega_1$ and $\omega_2$ are the frequencies of photons in a pair which add up to the frequency of the pumping laser $2\omega_0$, whereas $|\phi(\omega_1, \omega_2)|^2$ is a two-dimensional joint spectral profile of the photon pairs, and $f(\omega)$ is an object's transfer function which describes the phase delays which the object imparts on the light.

In both quantum and classical OCT methods, $f(\omega)$ is responsible for the appearance of fringes in the signal. In the Fd-Q-OCT signal, the transfer function appears in two different forms, each contributing to different elements in the A-scan after Fourier transformation. The term $|f(\omega_1)|^2 + |f(\omega_2)|^2$ will generate stationary artefact peaks. The height of these peaks will vary with $\omega_0$ and they will be located at a fixed distance from zero optical path difference (OPD), which corresponds to the zero point of the abscissa axis of the A-scan. For each artefact of this type, its distance will be equal to the distance between a pair of interfaces or scattering centres in the object this artefact is related to. Interestingly, since the Td-Q-OCT signal is an integration of (1) over $\omega_1$ and $\omega_2$, the term $|f(\omega_1)|^2 + |f(\omega_2)|^2$ averages out to a constant contribution and, consequently, does not generate this type of artefact in Td-Q-OCT. The last term in expression (1), $f(\omega_1)f^*(\omega_2)$, will lead to peaks representing object's dispersion-cancelled and resolution-doubled structure as well as a new type of an artefact: an instationary one which appears midway between two interfaces. This type of artefact is also present in Td-Q-OCT signals.

In the method of ICA-SD-OCT by Jensen *et al.*, a standard Fd-OCT signal is first Hilbert transformed so that its complex representation, $I_{\text{OCT}}$, is created. It is then used in a simple transformation to produce a two-dimensional output:

$$I_{ICA-SD-OCT}(\omega_1, \omega_2) = \text{Re}\{I_{OCT}(\omega_1)I_{OCT}^*(\omega_2)\} \tag{2}$$

Algorithmically, $I_{OCT}(\omega_1)$ is the complex spectrum and $I_{OCT}^*(\omega_2)$ corresponds to a reversed complex-conjugated spectrum. If we assume that the intensity of light coming back from the object and reference arms is equal, then $I_{OCT}(\omega)$ can be written as:

$$I_{OCT}(\omega) = 2I_0(\omega)|1 + f(\omega)|^2 = 2I_0(\omega)\big(1 + \text{Re}\{f(\omega)\} + |f(\omega)|^2\big)^2, \tag{3}$$

where $I_0$ is the source's spectrum. In (3), $\text{Re}\{f(\omega)\}$ is a cross-correlation term representing the depth structure of the object, and $|f(\omega)|^2$ is an auto-correlation term arising due to the interference of light backscattered from different parts of the object and responsible for peaks located close to 0 OPD in the A-scan. Using (3) in (2) gives:

$$I_{ICA-SD-OCT}(\omega_1, \omega_2) = |I_0(\omega)|^2(|f(\omega_1)|^2 + |f(\omega_2)|^2 + 2\,\text{Re}\{f(\omega_1)f^*(\omega_2)\}$$
$$+2\text{Re}\{f(\omega_1) + f(\omega_2)\} + f(\omega_1)|f(\omega_2)|^2 + f(\omega_2)|f(\omega_1)|^2 + |f(\omega_1)|^2|f(\omega_2)|^2 + 1) \tag{4}$$

The ICA-SD-OCT signal incorporates exactly the same terms as the Fd-Q-OCT signal: $|f(\omega_1)|^2 + |f(\omega_2)|^2$ responsible for the appearance of the stationary artefacts, and $f(\omega_1)f^*(\omega_2)$, which produces dispersion-cancelled and resolution-doubled structure positioned at twice the distance from 0 OPD as well as the instationary artefacts. In the ICA-SD-OCT signal, there are additional four terms, which give rise to more artefacts in the A-scan. The term $2\text{Re}\{f(\omega_1) + f(\omega_2)\}$ will recreate the structure of an object at a standard distance from 0 OPD, but the peaks will have an oscillatory character of an artefact; $f(\omega_1)|f(\omega_2)|^2$ and $f(\omega_2)|f(\omega_1)|^2$ will be additional artefacts. As it will turn out later, this type of artefact will nearly overlap one other type. Finally, $|f(\omega_1)|^2|f(\omega_2)|^2$ is an auto-correlation term – a dispersion-cancelled and resolution-doubled equivalent of the auto-correlation term in traditional OCT ($|f(\omega)|^2$ in (3)). In summary, in Td-Q-OCT, there is one artefact per pair of interfaces/scattering centres, in Fd-Q-OCT – two artefacts per pair, and in ICA-SD-OCT, there are three artefacts per pair plus additional $n$ artefacts ($n$ – number of interfaces/scattering centres in the object).

Apart from an increased number of artefacts, ICA-SD-OCT may also seem less attractive in terms of the axial resolution improvement. Although the resolution-doubling term $f(\omega_1)f^*(\omega_2)$ is present in both the Q-OCT signal and ICA-SD-OCT signal, the axial resolution of the latter is only improved by a factor of $\sqrt{2}$. As showed by Shirai and Friberg[24], this is due to the fact that to get the quantum-mimic OCT signal, the spectral profile, $I_0$, needs to be squared as well, as seen in equation (4)), which in the case of Gaussian-like profiles reduces the profile's width – and consequently the resolution – by $\sqrt{2}$. An obvious way around this problem is to use light sources with square-like spectral profiles which are transformed with little to no loss in full width at half maximum.

**Signals in Q-OCT and ICA-SD-OCT for an object with two scattering centres**

To see where exactly the resolution increase comes from in both quantum and quantum-mimic OCT, we first derive explicit expressions for their signals in the simplest case of two scattering centres as an object. We adapt the notation from the paper of Jensen *et al.* and write the formula expressing transfer function of an object consisting of two scattering centres in the following form:

$$f(\omega) = \text{Re}\left\{r_1 e^{\frac{\omega\Delta l}{c} + \beta(\omega)L_1} + r_2 e^{\frac{\omega\Delta l}{c} + \beta(\omega)L_2}\right\} \tag{5}$$

where $r_1$, $r_2$ are complex reflection coefficients, $\Delta L = L_2 - L_1$ with $L_1$, $L_2$ being twice the distances from the sample's surface to each of the scattering centres, $\Delta l = l_s - l_r$ with $l_s$, $l_r$ being the object and reference paths measured as twice the distance from the beam-splitter to the sample surface and reference mirror, respectively (see the schematics in Fig. 1 for more clarity), $\beta(\omega) = \frac{n(\omega)\omega}{c}$ is the wavenumber, with $c$ being the vacuum speed of light, and $n$ being the depth-averaged refractive index of the sample. This depth-averaged refractive index is different for two scattering centres at different depths, but this difference is assumed to be negligible for the clarity of the calculations. To account for higher order dispersion terms, $\beta = \beta(\omega)$ is expanded into Taylor series: $\beta(\omega) = \sum_{j=1} \frac{\beta_j \omega^j}{j!} = \beta_0 + \beta_1 \omega' + \beta_{NL}^{(even)} + \beta_{NL}^{(odd)}$, where $\beta_{NL}^{(even)} = \sum_{i=1} \frac{\beta_{2i}\omega'^{2i}}{2i!}$ and $\beta_{NL}^{(odd)} = \sum_{i=1} \frac{\beta_{2i+1}\omega'^{2i+1}}{(2i+1)!}$ are even and odd-order dispersion terms.

We substitute (5) in (1) with $\omega_1 = \omega_0 + \omega'$ and $\omega_2 = \omega_0 - \omega'$, where $\omega_0$ is a central frequency which "identifies" a diagonal in the two-dimensional joint spectrum and $\omega'$ is a frequency shift from the central frequency along the diagonal. Experimentally, a bigger number of diagonals is ensured by a spectrally broad pump laser. The substitutions give an explicit expression for the Fd-Q-OCT signal in the case of two scattering centres as an object:

$$C_{Fd-Q-OCT}(\omega_0, \omega') = \mathrm{Re}\left\{2r_1^2 e^{i\left[2\frac{\omega'}{c}\Delta l + 2L_1\left(\beta_1\omega' + \beta_{NL}^{(odd)}\right)\right]} + 2r_2^2 e^{i\left[2\frac{\omega'}{c}\Delta l + 2L_2\left(\beta_1\omega' + \beta_{NL}^{(odd)}\right)\right]}\right\} +$$
$$\mathrm{Re}\left\{4r_1r_2 e^{i\left[2\frac{\omega'}{c}\Delta l + (L_1+L_2)\left(\beta_1\omega' + \beta_{NL}^{(even)} + \beta_{NL}^{(odd)}\right)\right]}\right\}\cos(\Delta L\beta_0) +$$
$$\mathrm{Re}\left\{4r_1r_2 e^{i\left[\Delta L\left(\beta_1\omega' + \beta_{NL}^{(even)} + \beta_{NL}^{(odd)}\right)\right]}\right\}\cos(\Delta L\beta_0) + 2(r_1^2 + r_2^2)$$

(6)

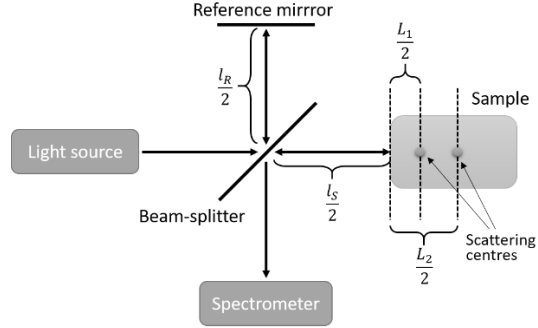

**Figure 1.** A schematic drawing of an interferometer with path lengths used for calculations.

The terms in the first line corresponds to the dispersion-cancelled structure shifted to twice the distance from 0 OPD due to the multiplication by a factor of 2. Also, what can be clearly seen in (6) – and what was not showed explicitly by authors of the paper introducing Fd-Q-OCT[16] – is that the instationary and stationary artefacts (described by the terms in the second and third lines) oscillate with $\omega_0$ at the same rate: $\cos(\Delta L\beta_0 = \cos\left(\Delta L\frac{n(\omega_0)}{c}\omega_0\right)$ with $\Delta L\frac{n(\omega_0)}{c}$ being the frequency. As expected, the artefacts incorporate both even- and odd-order dispersion terms.

(5) is used in (4), again with $\omega_1 = \omega_0 + \omega'$ and $\omega_2 = \omega_0 - \omega'$. In this case, $\omega_0$ is the central frequency of an OCT spectrum and $\omega'$ is the frequency shift from the central frequency. This gives an explicit expression for the ICA-SD-OCT signal in the case of two scattering centres as an object (repeated from the paper of Jensen et al., but with several corrections):

$$I_{ICA-SD-OCT}(\omega_0, \omega') = \mathrm{Re}\left\{4r_1^2 e^{i\left[2\frac{\omega'}{c}\Delta l + 2L_1\left(\beta_1\omega' + \beta_{NL}^{(odd)}\right)\right]} + 4r_2^2 e^{i\left[2\frac{\omega'}{c}\Delta l + 2L_2\left(\beta_1\omega' + \beta_{NL}^{(odd)}\right)\right]} + 4r_1^2r_2^2 e^{i\left[2\Delta L\left(\beta_1\omega' + \beta_{NL}^{(odd)}\right)\right]}\right\} +$$
$$\mathrm{Re}\left\{8r_1r_2 e^{i\left[2\frac{\omega'}{c}\Delta l + (L_1+L_2)\left(\beta_1\omega' + \beta_{NL}^{(even)} + \beta_{NL}^{(odd)}\right)\right]}\right\}\cos(\Delta L\beta_0) +$$
$$\mathrm{Re}\left\{4r_1r_2(r_1^2 + r_2^2)e^{i\left[\Delta L\left(\beta_1\omega' + \beta_{NL}^{(even)} + \beta_{NL}^{(odd)}\right)\right]}\right\}\cos(\Delta L\beta_0) +$$
$$\mathrm{Re}\left\{4r_1(r_1^2 + r_2^2)e^{i\left[\frac{\omega'}{c}\Delta l + L_1\left(\beta_1\omega' + \beta_{NL}^{(even)} + \beta_{NL}^{(odd)}\right)\right]}\right\}\cos\left(\Delta l\frac{\omega_0}{c} + L_1\beta_0\right) +$$
$$\mathrm{Re}\left\{4r_2(r_1^2 + r_2^2)e^{i\left[\frac{\omega'}{c}\Delta l + L_2\left(\beta_1\omega' + \beta_{NL}^{(even)} + \beta_{NL}^{(odd)}\right)\right]}\right\}\cos\left(\Delta l\frac{\omega_0}{c} + L_2\beta_0\right) +$$
$$\mathrm{Re}\left\{2r_1r_2^2 e^{i\left[\frac{\omega'}{c}\Delta l + 2L_2\left(\beta_1\omega' + \beta_{NL}^{(odd)}\right) + L_1\left(\beta_1\omega' + \beta_{NL}^{(even)} + \beta_{NL}^{(odd)}\right)\right]}\right\}\cos\left(\Delta l\frac{\omega_0}{c} + L_1\beta_0\right) +$$
$$\mathrm{Re}\left\{2r_1^2r_2 e^{i\left[\frac{\omega'}{c}\Delta l + 2L_2\left(\beta_1\omega' + \beta_{NL}^{(even)} + \beta_{NL}^{(odd)}\right) + 2L_1\beta_{NL}^{(even)}\right]}\right\}\cos\left(\Delta l\frac{\omega_0}{c} + (\Delta L - L_1)\beta_0\right) + 2(r_1^2 + r_2^2)$$

(7)

The first two terms in line 1 correspond to cross-correlation peaks representing the structure of the object, and the third one – to the auto-correlation peak, all moved to twice the distance from 0 OPD in the A-scan due to the multiplication by a factor of 2 (the first scatterer placed at the distance $\sim (2\Delta l + 2L_1)$, the second scatterer at $\sim (2\Delta l + 2L_2)$ and the auto-correlation peak at $\sim 2\Delta L$, respectively). The terms in lines 2 and 3 are responsible for the instationary and stationary artefacts, the next ones in lines 4 and 5 – for artefacts recreating the structure at standard distances ($\sim (\Delta l + L1)$ and $\sim (\Delta l + L2)$), one in line 6 represents an artefact peak placed at a distance $\sim (\Delta l + \Delta L + L2)$ and the first one in the last line – an artefact placed around the location of an artefact expressed as the term in line 5.

To visualise the similarities between the two methods, imaging of an object consisting of two scattering centres was simulated for Fd-OCT and Fd-Q-OCT systems with similar optical parameters: central wavelength of 1560 nm and the total spectral bandwidth of 115 nm. The scatterers were placed 100 $\mu m$ apart and 0.5 mm away from 0 OPD in the A-scan. We assumed the refractive index, group refractive index and group velocity dispersion ($\beta_2$) to be of quartz at 1560 nm.

An OCT spectrum (Fig. 2a) was processed using the ICA-SD-OCT algorithm in its original form: the spectrum was first split into smaller fragments – each 86 nm wide and centred at 200 different wavelengths within the range 1546 -1557 nm (these central wavelengths correspond to $\omega_0$). Every fragment was transformed using (2) and put one on top of another to create a spectral stack depicted in Fig. 2c. In the case of Fd-Q-OCT, diagonals of a joint spectrum were calculated and also put one on top another to create a spectral stack (Fig. 2e). All rows from each such stack were Fourier transformed to create corresponding FFT stacks (Fig. 2d and f). As expected, the FFT stack in ICA-SD-OCT contains the same elements as the FFT stack in Fd-Q-OCT as well as some additional artefacts. The depth axes were not recalculated to optical distance to visualise that the positions of the peaks are moved to twice the distance from 0 OPD in an A-scan in Quantum and quantum-mimic OCT (Fig. 2d and f) compared to traditional OCT (Fig. 2b).

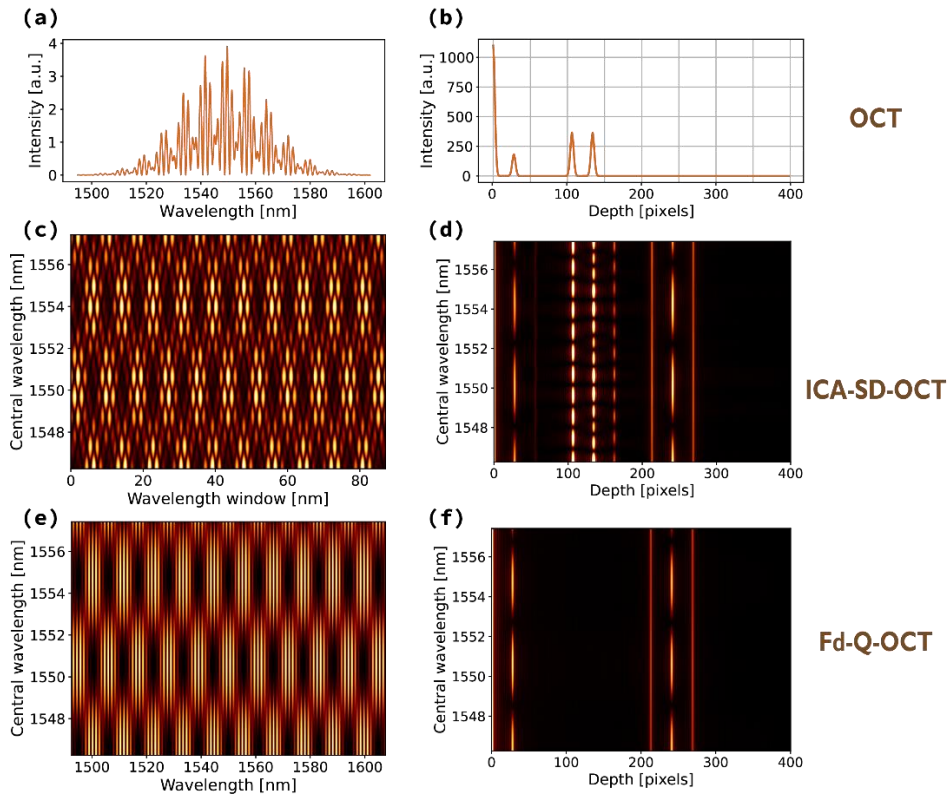

**Figure 2. (a)** A simulated spectrum in Fd-OCT corresponding to an object consisting of two scattering centres. **(b)** Its Fourier transform depicting two structural peaks representing the object (cross-correlation terms) and a peak around 0 which appears due to the interference of light back-scattered by the object (auto-correlation term). **(c)** Consecutive fragments of an Fd-OCT spectrum processed with the ICA-SD-OCT algorithm put one on top of another in a spectral stack, **(d)** Fourier transformed rows of the spectral stack show structural peaks at twice the distance from 0 OPD, an auto-correlation term and five varying artefacts. **(e)** Diagonals of a joint spectrum put one on top of another in a spectral stack, **(f)** Fourier transformed rows of the spectral stack show structural peaks at twice the distance from 0 OPD and two artefacts

**Twofold axial resolution improvement**

The key parameter enabling axial resolution improvement in ICA-SD-OCT, or any of the quantum-mimic OCT methods that are based on the principle of spectral intensity correlation, is the factor of 2 by which the phase arguments of all the structural components in the signals are multiplied. We see this multiplication in (7) describing the quantum-mimic signal as well as in (6) describing an Fd-Q-OCT signal. In both quantum-mimic and quantum OCT, the multiplication by 2 doubles the frequency of all the oscillatory components of a corresponding spectrum. Since the frequency of these components is in direct relation to the position of the structural peaks in the A-scan, twofold frequency increase will lead to a twofold displacement of the peaks in an A-scan and, as a result, to a twofold increase of the distance between the two peaks. This means that if in a standard A-scan two peaks are placed so close to each other that they start to overlap completely and are therefore no longer distinguishable, application of a quantum-mimic or quantum method will spread these peaks apart and can make them distinguishable in the resulting A-scan.

This matter can be looked at from the point of view of an axial resolution formula $\delta z = \frac{2\ln 2}{\pi} \frac{1}{\Delta k}$. An oscillatory component, $f$, of an OCT signal is in practice a sum of cosine functions of $R\cos(Zk)$ form, where $R$ is the reflection coefficient and $Z$ is twice the distance between the scatterers in the object arm. The variable $k$ varies between $k_1$ and $k_2$, which represents a total spectral bandwidth of a light source with a Gaussian spectrum whose full width at half maximum is $\Delta k$. In quantum-mimic and quantum regime, the same component is written as $R\cos(2Zk)$. The factor of 2 in the cosine can be "grouped" with $Z$ to explain the twofold frequency increase of the cosine as in the previous paragraph. Alternatively, the factor of 2 could be grouped with $k$. In such a case, a multiplication of $k$ by 2 will lead to "moving" the limits of the spectral bandwidth to $2k_1$ and $2k_2$, so consequently to doubling of the spectral bandwidth, and with it, to doubling of $\Delta k$, which will directly reflect on the axial resolution: $\delta z_{\text{quantum}} = \frac{2\ln 2}{\pi} \frac{1}{2\Delta k} = \frac{1}{2}\left(\frac{2\ln 2}{\pi} \frac{1}{\Delta k}\right) = \frac{1}{2}\delta z$.

An experimental verification of this phenomenon is challenging due to the presence of artefacts. To eliminate them, Jensen *et al.* used the fact that the artefacts' height changes as a function of the central frequency $\omega_0$ (expressed by the cosine components of all the artefact terms in (7)). As a result, a mean of the spectral fragments – each centred at a different $\omega_0$ – leads to reduction of artefacts. It was shown that this reduction is the most effective when the spectral fragments allow for at least five oscillations of an artefact. Interestingly, this algorithm can be treated as a numerical equivalent of an experimental artefact suppression used by Graciano *et al.*[31] in Td-Q-OCT. In their system, they used a broadband pump laser to generate a joint spectrum which is broad in the anti-diagonal direction. A larger anti-diagonal width translates to more diagonals, each centred at a different $\omega_0$. Performing a time-domain detection leads to a coherent averaging of all the diagonals and consequently, reduction of the artefacts.

The period of the artefacts' oscillation depends on the distance between the peaks: the smaller the distance, the bigger the period (see Fig. 4a-c depicting FFT stacks for single-layer objects with varying thicknesses). It makes the suppression of artefacts very challenging for objects whose thickness lies at the axial resolution limit, because too closely spaced scatterers create artefacts with periods requiring an $\omega_0$ span too big for the artefact suppression algorithm to be successful. Consequently, although the twofold resolution enhancement is present in quantum-mimic (and quantum) OCT methods, it can be obscured by the artefacts, more specifically mainly by the instationary artefact which is always located midway between every two peaks.

## Experimental results

**Change of axial resolution with the spectral shape**

To show how the spectral profile's shape affects the axial resolution, two interference signals were measured in an Fd-OCT system using light centred at 1560 nm (total spectral bandwidth of 115 nm) and a mirror as an object. The first signal was shaped to present a Gaussian-like profile (Fig. 3a) and the other one – to have a square-like profile (Fig. 3b). Both spectra were processed according to (2). The absolute value of the output, an ICA-SD-OCT spectrum, is plotted in Fig. 3c and d. Finally, both ICA-SD-OCT spectra were Fourier transformed and the absolute value of the outputs were plotted as A-scans in Fig. 3e and f in a light brown colour. For comparison, the original interference spectra were also Fourier transformed and plotted on the same graphs in a dark brown colour.

The insets in the graphs in Fig. 3e and f show an area around the peaks that represent the position of the mirror. The width of the peaks in the A-scans obtained with the original OCT spectra is approximately 22 $\mu m$. In theory, a spectrum with a square-like shape provides a better axial resolution than a Gaussian-shaped spectrum in the same spectral range. In our measurements, we performed a coarse hardware spectral shaping which resulted in narrowing of the effective wavelength range of both spectra (and a shift of central wavelength). The extent of narrowing was bigger for the square-like spectrum which led to decreasing of axial resolution down to the level of the axial resolution corresponding to the Gaussian-like spectrum.

In the case of the Gaussian-like profile, using the ICA-SD-OCT algorithm decreased the width of the peak to 14 $\mu m$ which translates to around 22/14=1.57 axial resolution increase. The increase is bigger than the theoretical 1.41, because the spectrum's shape was not a perfect Gaussian. In the case of the square-like spectrum, the width dropped to 12 $\mu m$ giving over 1.8-fold axial resolution improvement. We note that the axial resolution increase would be even higher if the spectrum was shaped into a square function shape in a more precise way.

Fig. 3e and f show one artefact in the ICA-SD-OCT A-scans (light brown line). Indeed, when $r_2 = 0$, in the equation (7) describing an ICA-SD-OCT signal for two scatterers only two terms are left: one corresponding to the structural peak and the other – to the artefact placed at half the location of the structural peak. Also Fig. 3e and f show how unbalanced dispersion affects the artefacts in ICA-SD-OCT A-scans. Whereas in Fig. 3e the dispersion mismatch was very small, in Fig. 3f, it was substantial due to an additional piece of glass that was present in the reference arm at the time of the measurement. Since numerical dispersion compensation does not affect artefacts and even-order dispersion terms are inherently not cancelled in their case (see (7)), the artefact in Fig. 3f remains broadened and distorted by higher-order dispersion introduced by the excess amount of glass in the reference arm. On the other hand, the DC peak is affected by numerical dispersion compensation. The DC peak is related to the spectral profile which is not changed by unbalanced dispersion in the interferometer. Due to numerical dispersion compensation – which is a multiplication of the spectrum by a complex phase factor incorporating both linear and nonlinear terms negating the impact of unbalanced dispersion – the DC peak is shifted and broadened in an OCT A-scan (dark brown line in Fig. 3f), but only shifted in the ICA-SD-OCT A-scan (light brown line in Fig. 3f), because dominant second-order dispersion is automatically cancelled. In the case of the structural peak in the ICA-SD-OCT A-scan in Fig. 3f, numerical dispersion compensation was only applied to compensate for the third-order dispersion.

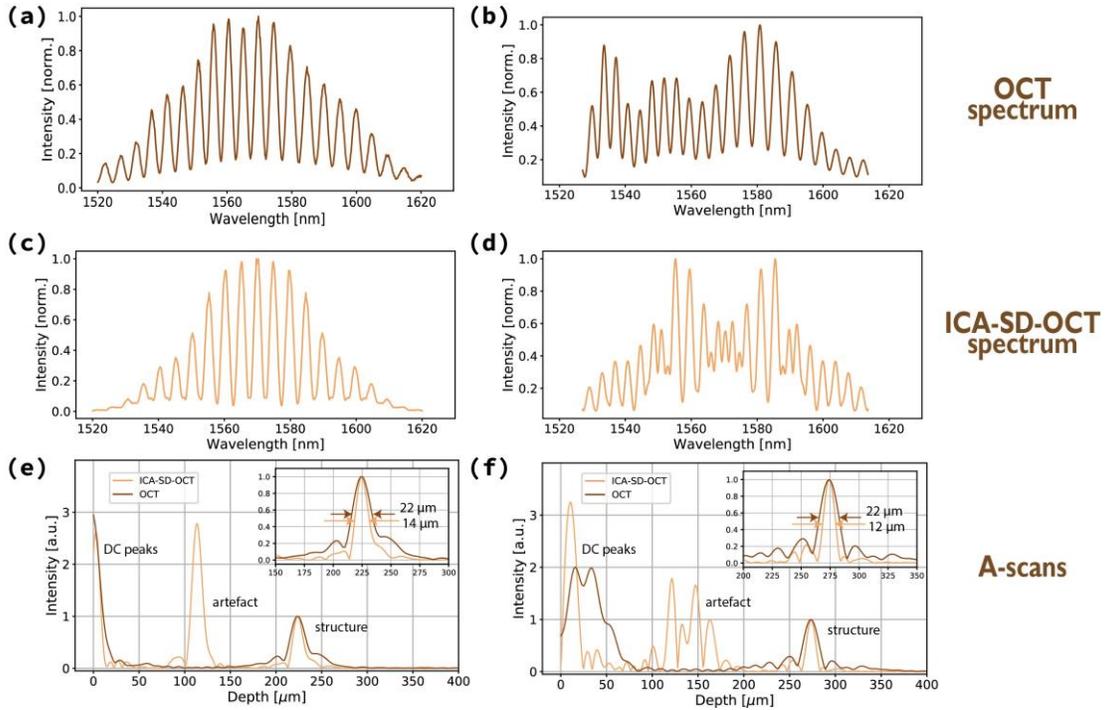

**Figure 3.** Experimentally measured Gaussian-like spectrum **(a)** and square-like spectrum **(b)**. The results of transformation according to formula (2) depicted in graphs **(c)** and **(d)**, were Fourier transformed to produce A-scans (light brown in **e** and **f**) with a better axial resolution than A-scans corresponding to the original OCT spectra (dark brown in **e** and **f**). The insets with a zoom-in on the structural peaks show a 1.57 axial resolution increase in the case of the Gaussian-like profile, and a 1.83 increase in the case of the square-like profile. A difference in the shape of DC peaks and artefacts between (e) and (f) are due to different dispersion imbalance in the interferometer at the time of measurements.

**Axial resolution enhancement**

Fig. 4a-c present FFT stacks for 0.26-mm thick BK7, 0.14-mm thick sapphire and 0.08-mm thick quartz. For the quartz, the period of the instationary artefact becomes very big and, due to the proximity with the structural peak, this artefact starts to overlap the structure and interferes with it. Hence, estimation of the axial resolution improvement is challenging. To provide an experimental proof that ICA-SD-OCT is able to resolve two very closely spaced scatterers in a situation when traditional OCT cannot, we decided to use a spectrum for quartz and compared an A-scan calculated using a spectrum which was a 20-nm wide fragment of an original spectrum with an A-scan obtained using the ICA-SD-OCT algorithm on enough 20-nm spectral fragments to suppress the instationary artefact peak. Whereas in the first A-scan (dark brown line in Fig. 4d) the structure of the quartz is not resolved, in the second one (plotted in light brown line in Fig. 4d), it is. An FFT stack obtained by Fourier transforming the ICA-SD-OCT-processed spectral fragments is presented in Fig. 4e and shows that one full well-sampled oscillation of an artefact was enough to suppress it completely.

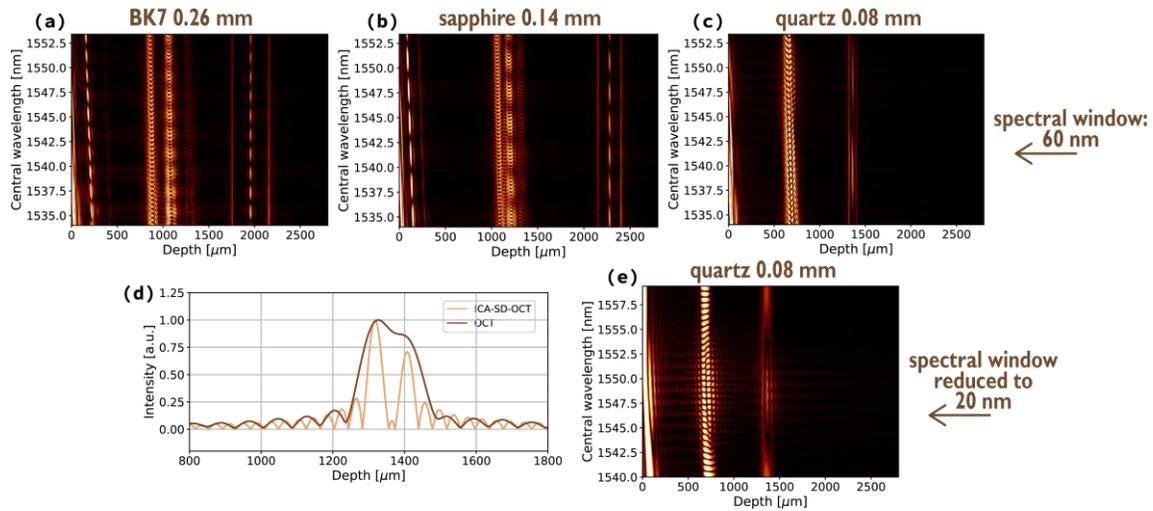

**Figure 4.** FFT stacks obtained with 60-nm wide fragments of an OCT spectrum for 0.26-mm thick BK7 (a), 0.14-mm thick sapphire (b) and 0.08-mm thick quartz (c) show that the period of the artefacts increases with a decreasing thickness. (d) The quartz structure is not resolved using a 20 nm wide spectrum (dark brown line), but becomes resolved when the artefacts are removed after averaging several 20-nm wide spectra (light brown line). (e) FFT stack of quartz corresponding to a reduced width of an individual spectrum.

Also, two interesting effects can be observed which occur due to numerical dispersion compensation. First, as it was already discussed in the previous subsection, the artefact peaks get distorted. Second, the position of the artefacts changes with varying central wavelength (corresponding to varying $\omega_0$). It is due to the fact that numerical dispersion compensation does not affect the artefacts and leaves dispersion-induced non-linearity in the cosine functions they are represented with. As a result, different fragments of a spectrum will correspond to a different frequency and therefore to a different position of a peak.

## Discussion and outlook

ICA-SD-OCT algorithm processes an OCT spectrum into what can be interpreted as diagonals of Fd-Q-OCT's joint spectrum and consequently, reproduces all the advantageous aspects of the Q-OCT signal: axial resolution enhancement and dispersion cancellation. On the other hand, it also recreates Q-OCT's artefacts and, on top of that, produces several additional artefacts. An increased number of artefacts in quantum-mimic OCT may seem disadvantageous as compared to Q-OCT, but from the viewpoint of biomedical imaging, which requires a perfect correspondence of the image to the imaged object, a single artefact is still one too many.

What works to the advantage of quantum-mimic OCT is the fact that it gives very easy access to features that were thought to be only possible through very expensive and time-consuming experiments in quantum optics. By being a simple algorithm, it can potentially be used with existing Fd-OCT systems to easily enhance their quality and resolution. The only remaining roadblock on the path to achieve this are artefacts. There are already both hardware and software schemes to reduce the artefacts and, in some well-defined cases, suppress them completely, but they are yet to be universal to work in practical OCT imaging scenarios.

Because the behaviour of artefacts is the same in quantum OCT and quantum-mimic OCT, the schemes for artefact removal in quantum OCT are easily adapted for use in quantum-mimic OCT (and vice versa). Such close relationship of quantum and quantum-mimic OCT on this and other levels creates a unique opportunity for very original solutions since it makes room for research methods from two fundamentally different sides of optics: classical and quantum. Such joint forces will undoubtedly enable the creation of methods which will remove the remaining obstacles and lead to substantial advancement of OCT imaging.

## Methods

The OCT system used for the experiments is presented in Fig. 5. It is the same setup used in our previous work[32] and here we will repeat its main characteristics.

Pulsed light with a central wavelength of 1560 nm and a total spectral bandwidth of 115 nm (MenloSystems T-Light) is inputted into a Linnik-Michelson interferometer through a fibre collimator FB1 (f=11 mm). The repetition rate of the laser, 100 MHz, allows for a temporal broadening of up to 10 ns before adjacent pulses start to overlap each other. In the detection part, we used a 5-km long fibre spool (SMF28E, Fibrain) with a group velocity dispersion, $\beta_2$, equal to 23 fs$^2$/mm which broadened the pulses coming from the interferometer to 9.6 ns. The output port of the fiber spool was monitored by a Superconducting Single-Photon Detector (SSPD) (Scontel) with a detection range is 350-2300 nm and a peak quantum efficiency of approximately 65% at 1550 nm[33]. The SSPD outputs an electric pulse after each successful detection of a single photon and the FPGA electronics measures the timestamps of the electric pulses. The timing jitter of the apparatus consisting of the SSPD and the time tagging unit is 35 ps, which is very close to the state of the art, but an order of magnitude worse than for a standard photodiode. Due to the high sensitivity of the SSPD, the light source was attenuated to a level of single photons per pulse by using a half-wave plate and a polarization beam-splitter at the input of the interferometer. The SSPD was synchronized with the fast built-in photodiode in the light source. Because the fibre spool – through the phenomenon of dispersion – delays each wavelength by a different amount of time, time measurement performed by the SSPD provided a spectrum of the light at the input of the fibre collimator FB2 (f=11 mm). Measured spectra were digitized by FPGA electronics and saved onto a computer. Because the fibre spool's dispersion curve is not a linear function in wavenumber, a linearisation of the acquired spectra was performed[7].

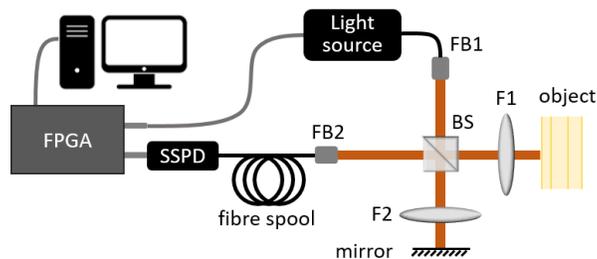

**Figure 5.** Experimental setup (reproduced from Ref.[32]). The light source is a pulsed laser attenuated to a level of single photon per pulse. Pulses are coupled to a fibre (FB1) and propagate in a Linnik-Michelson interferometer. The input wavepacket (pulse) is then split at a beamsplitter (BS) into two arms. In the object arm, one wavepacket interacts with the object and acquires an additional phase; in the reference arm, the other one is reflected from the mirror. They both overlap at the beamsplitter and the output is coupled to a single-mode fibre spool using a fibre coupler FB2. The time-resolving Superconducting Single-Photon Detector (SSPD) together with the long dispersive fibre spool work as a spectrometer. Time reference is provided by a photodiode signal from the light source. The data is collected using an FPGA time-stamping electronics. F1, F2 – lenses.

A coarse hardware spectral shaping discussed around Fig. 3 was performed using the half-wave plate and the polarization beam-splitter placed at the input of the interferometer. Rotation of the half-wave plate in front of the polarization beam-splitter led to simultaneous decrease in intensity and change in spectral shape of the spectrum enabling two spectral profiles: Gaussian and square-like.

## Acknowledgements

SMK acknowledges The Dodd-Walls Centre for Photonic and Quantum Technologies (New Ideas Fund). PK acknowledges financial support by the Foundation for Polish Science (FNP) (project First Team co-financed by the European Union under the European Regional Development Fund).

## Author contributions statement

S.M.K. conceived the experiments, conducted them in P.K.'s laboratory and analysed the results. All authors reviewed the manuscript.